\documentclass[fleqn,usenatbib]{mnras}

\usepackage {psfig}
\usepackage{amssymb}
\usepackage {graphicx}
\usepackage {epsfig}
\usepackage{mathrsfs}
\usepackage[english]{babel}
\usepackage{amsfonts}
\usepackage{fontenc}
\usepackage{textcomp}
\usepackage{microtype}
\usepackage[pdftex]{lscape}
\usepackage{amsmath}

\title[Non-thermal states in filaments]{Dynamical origin of non-thermal states in galactic filaments}
\author[Di Cintio, Gupta and Casetti]{Pierfrancesco Di Cintio$^{1,2}$\thanks{E-mail:p.dicintio@ifac.cnr.it}, Shamik Gupta$^{3}$\thanks{E-mail: shamikg1@gmail.com}, and Lapo Casetti$^{2,4,5}$\thanks{E-mail:lapo.casetti@unifi.it}\\
$^{1}$Consiglio Nazionale delle Ricerche, Istituto di Fisica Applicata ``Nello Carrara",\\
 via Madonna del piano 10, I-50019 Sesto Fiorentino, Italy\\
$^{2}$INFN -  Sezione di Firenze, via G.\ Sansone 1, I-50019 Sesto Fiorentino, Italy\\
$^{3}$Department of Physics, Ramakrishna Mission Vivekananda University, Belur Math, Howrah 711202 West Bengal, India\\
$^4$Dipartimento di Fisica e Astronomia and CSDC, Universit\`a di Firenze, via G.\ Sansone 1, I-50019 Sesto Fiorentino, Italy\\
$^5$ INAF - Osservatorio Astrofisico di Arcetri, largo Enrico Fermi 5, I-50125 Firenze, Italy}
\begin{document}
\date{Accepted...  Received...; in original form...}
\pagerange{\pageref{firstpage}--\pageref{lastpage}} 
\pubyear{0000}
\maketitle

\begin{abstract}
Observations strongly suggest that filaments in galactic molecular clouds are in a non-thermal state. As a simple model of a filament we study a two-dimensional system of self-gravitating point particles by means of numerical simulations of the dynamics, with various methods: direct $N$-body integration of the equations of motion, particle-in-cell simulations and a recently developed numerical scheme that includes multiparticle collisions in a particle-in-cell approach. Studying the collapse of Gaussian overdensities we find that after the damping of virial oscillations the system settles in a non-thermal steady state whose radial density profile is similar to the observed ones, thus suggesting a dynamical origin of the non-thermal states observed in real filaments. Moreover, for sufficiently cold collapses the density profiles are anticorrelated with the kinetic temperature, i.e., exhibit temperature inversion, again a feature that has been found in some observations of filaments. 
The same happens in the state reached after a strong perturbation of an initially isothermal cylinder. Finally, we discuss our results in the light of recent findings in other contexts (including non-astrophysical ones) and argue that the same kind of non-thermal states may be observed in any physical system with long-range interactions. \end{abstract}


\section{Introduction}
Many astrophysical systems are found in states out of thermal equilibrium. This is not surprising especially when gravity is the dominant interaction, given the overwhelmingly large relaxation times\footnote{Strictly speaking, ``true'' thermodynamic equilibrium is impossible in isolated and finite self-gravitating systems in three dimensions, due to the non-confining nature of the potential; in this case the relaxation time denotes the time needed to effectively erase the memory of the initial condition. However, in the following we shall be mainly concerned with systems that are effectively two-dimensional, where the gravitational potential becomes confining and a thermal equilibrium state is possible.} that grow with the number of constituents of a self-gravitating system (see e.g.\ \citealt{2008gady.book.....B}).
Recent observations, especially those obtained in submillimetre band by the {\em Herschel Space Observatory}, revealed that filaments in molecular clouds are not an exception to this rule, i.e., exhibit density profiles that are not consistent with thermal equilibrium (\citealt{2011A&A...529L...6A}; see also the discussion in \citealt{2015MNRAS.446.2110T}). 
Filaments are elongated structures present in galactic molecular clouds with densities ranging from $10^4$ to $10^6$ particles per ${\rm cm^3}$ that extend for tens of parsecs in length and that typically host star forming cores (see e.g.\ \citealt{2011EAS....51..133K,2015ApJ...806..226M}). Molecular filaments are expected to be short-lived ($\approx 10^7$ yr) and, at least in their initial stages, gravitationally supported structures (\citealt{2014ApJ...791..124G}, see also \citealt{2004ApJ...616..288B}) embedded in galactic magnetic fields (\citealt{1996ApJ...472..673G,2012ARA&A..50...29C,2013MNRAS.436.3707L,2015MNRAS.446.2118T}, and references therein). Another intriguing feature of the non-thermal states of filaments is that the radial density $\rho(r)$ and kinetic temperature $T(r)$ profiles are often anticorrelated, i.e., the kinetic temperature of the gas is higher at larger radii where the density is smaller; this emerges indirectly, from the fact that density profiles are well fitted by polytropic solutions yielding $T$ versus $\rho$ power laws with negative exponents (\citealt{2015MNRAS.446.2110T}), and also from some direct observations of velocity dispersion (\citealt{0004-637X-504-1-223,2010ApJ...712L.116P}). This feature, dubbed ``temperature inversion'', is often found also in astrophysical plasmas encompassing a broad range of sizes and temperatures, from, for example, cool cores of galaxy clusters (\citealt{0004-637X-601-1-184}) to the hot ($5\times 10^6$ K) plasma torus surrounding Jupiter's moon Io (\citealt{1993JGR....9821163M,MeyerVernet1995202,2004jpsm.book..537S}). Stellar coronae also show temperature profiles increasing with radius (from $10^3$ K at the photosphere up to $10^6$ K for our Sun) where instead the density rapidly falls off with radius (see e.g.\  \citealt{2005psci.book.....A,GolubPasachoff:book}, and references therein). Observations have been reported of elliptical galaxies where the velocity dispersion increases with the distance from the nucleus \citep{LoubserEtAl:mnras2008,2017arXiv170800870V}. 
Transient anticorrelated density and temperature profiles have been observed also in one-dimensional self-gravitating systems \citep{W:thesis}. Typical non-thermal features that might bear a relation with temperature inversion are core--halo structures, that are ubiquitous in long-range-interacting systems \citep{LevinEtAlphysrep:2014}, having been observed in charged particle beams \citep{2014ApPhL.104g4109N,2001picp.book.....D,2008PhRvS..11h2801H}, in one-dimensional toy models of gravity \citep{1969MNRAS.146..161C,2011MNRAS.417L..21T,2013MNRAS.431...49S}, and in two- and three-dimensional models of self-gravitating systems \citep{2015MNRAS.451..622R,1988SvA....32..374D,1999ApJ...518..233D}. 
Recently it has been suggested that non-thermal states with temperature inversion may be typical of any long-range-interacting many-body system (\citealt{2014EPJB...87...91C,2015PhRvE..92b0101T,2016PhRvE..93f6102T}), being long-time quasi-stationary states of the dynamics arising after a violent relaxation process, and that could be observed also in condensed matter systems (\citealt{njp2016}).\\
\indent The observation of non-thermal states in filaments raised questions about the physical processes that may support such nonequilibrium states, and many ideas have been put forward, typically involving the r\^{o}le of local turbulence and of magnetic fields (see e.g.\ \citealt{2015MNRAS.446.2110T,2015MNRAS.446.2118T} and references therein). In this paper we argue that ingredients other than gravitational interactions may not be necessary at all, as such states may simply be the end states of the dynamics of the self-gravitating particles composing the filament, after a cold collapse as well as after a strong perturbation of an initially thermal state. To this end, we present a dynamical study of a very simple model of a filament, where both dissipative effects and magnetic fields are completely neglected, assuming our system is infinitely extended with perfect cylindrical symmetry, so that its dynamics reduces to that of a two-dimensional system of $N$ self-gravitating particles. We study the dynamics of the collapse of Gaussian overdensities, showing that after the damping of the virial oscillations the system settles in a quasi-stationary state whose density profile is well described by the empiric law used to model the radial density profile of observed filaments in \cite{2011A&A...529L...6A} and discussed by \cite{2015MNRAS.446.2110T} showing its relation to polytropic solutions of the fluid equations for the gas in the filament. For sufficiently cold collapses, the end states we find do exhibit temperature inversion. The same kind of non-thermal states with temperature inversion is observed after a strong perturbation of an initially thermal state.\\   
\indent
The paper is organized as follows. In Section \ref{sec:model} we first recall previously studied fluid models, then introduce our particle-based model and our numerical approach. In Section \ref{sec:collapse} we study the collapse of cylindrical Gaussian overdensities, with different initial values of the virial ratio and of the spread of the Gaussian, and in Section \ref{sec:perturb} we study the dynamics after a strong perturbation of an initial state in thermal equilibrium. In Section \ref{sec:discussion} we discuss our results and their connection with those found for other long-range-interacting many-body systems, also in non-astrophysical contexts, and finally in Section \ref{sec:conclusions} we draw our conclusions.

\section{Model and numerical techniques}
\label{sec:model}

\subsection{Fluid models}
\label{sec:fluidmodels}
Neglecting the presence of galactic magnetic fields, in the fluid picture the dynamical evolution of self-gravitating filaments is given in terms of mass density $\rho$, pressure $P$, gravitational potential $\Phi$ and velocity field $\mathbf{u}$ by 
\begin{equation}\label{fluideqs}
\begin{cases}
\displaystyle \partial_t\rho+\nabla\cdot(\rho\mathbf{u}) = 0\,,\\
\displaystyle \partial_t\mathbf{u} + (\mathbf{u}\cdot\nabla)\mathbf{u}=-\nabla\Phi-\frac{1}{\rho}\nabla P\,,\\
\displaystyle \Delta\Phi=4\pi G\rho\,,
\end{cases}
\end{equation}
where $G$ is the gravitational constant. Within the assumption of cylindrical symmetry, it is natural to express the operators in Eqs.\ (\ref{fluideqs}) in cylindrical coordinates $(r,z,\varphi)$ and to set $\partial_z \ = \partial_\varphi = 0$. Stationary solutions of Eqs.\  (\ref{fluideqs}) (i.e., with $\partial_t\rho = \partial_t \mathbf{u} = 0$ and\footnote{As noted by \cite{2015MNRAS.446.2110T}, stationary solutions with a nonzero velocity field are possible and might be relevant when modeling accretion phenomena or the so-called ``varicose'' instabilities.} $(\mathbf{u}\cdot\nabla)\mathbf{u} = 0$, and thus $\nabla\Phi=-\rho^{-1}\nabla P$) are typically obtained by imposing an equation of state relating pressure to density and solving for $\rho$ the differential equation obtained by combining the second and the third equations after having expressed $P$ as function of $\rho$. One of the most common forms used to model filamentary clouds is the polytropic equation of state (\citealt{1990ApJ...355..172R,1991ApJ...380..476A,2005MNRAS.356.1429S,2014MNRAS.437.2675B,2015MNRAS.454.2815L})
\begin{equation}
P=\kappa\rho^\gamma\,,
\end{equation}
where the constant $\kappa$ is connected to the system's entropy, and the so-called polytropic exponent $\gamma$ is related to the polytropic index $n$ by $\gamma=1+1/n$. In this case, the density profile supporting the polytropic equation of state is obtained by solving
\begin{equation}\label{polydiff}
\nabla\cdot(\rho^{-1}\nabla\rho^\gamma)+\frac{4\pi G}{\kappa}\rho=0\,,
\end{equation} 
that after exploiting the cylindrical symmetry and introducing dimensionless coordinates reduces to the standard cylindrical Lane-Emden equation (\citealt{2015MNRAS.446.2110T}).
In polytropic stationary models, the density profile is related to the temperature profile as $\rho\propto T^n$.\\
\indent A good approximation of the solution of Eq. (\ref{polydiff}) for infinite cylinders is given by the softened power-law radial density profile (\citealt{2015MNRAS.446.2110T,2016MNRAS.457..375F,2016A&A...592A..90C,2017arXiv170301394X})
\begin{equation}\label{softpower}
\rho(r)=\frac{\rho_c r_c^\alpha}{(r_c^2+r^2)^{\alpha/2}}\,,
\end{equation}
associated to the radial mass per unit length profile
\begin{equation}\label{massprof}
M_\ell(r)=2\pi\int_0^r\rho(r^\prime)r^\prime{\rm d}r^\prime=\frac{2\pi\rho_c r_c^\alpha}{(2-\alpha)}\left[\left(r^2+r_c^2\right)^{1-\alpha/2}-r_c^{2-\alpha}\right]\,.
\end{equation}
In the equations above $\rho_c$ and $r_c$ are the (finite) core density and core radius, respectively. Note that the total mass per unit length at height $z$, $M_\ell$, obtained extending to $+\infty$ the integral in Eq.\ (\ref{massprof}), is finite for $\alpha>2$. Note also that if $\alpha=4$ in Eq.\ (\ref{softpower}) we would obtain the density profile for the isothermal cylinder (\citealt{1963AcA....13...30S,1964ApJ...140.1056O,1964ApJ...140.1529O}). In this case the core radius $r_c$ becomes
\begin{equation}\label{rcostriker}
r_c=\sqrt{{2}/{\pi G\beta\rho_c}}\,,
\end{equation}
where $\beta=m_0/k_BT$ fixes the width of the equilibrium Maxwellian velocity distribution at temperature $T$ with mean molecular mass $m_0$, and $k_B$ is the Boltzmann constant. The parameter $\beta$ can be also expressed as function of the total mass\footnote{Note that $M_\ell$ is a finite quantity at variance with the diverging  mass of an isothermal sphere. The total mass of the isothermal filament, however, is indeed infinite, since the integral of $M_\ell$ over $z$ diverges.} per unit length $M_\ell$ as $\beta=2/GM_\ell$.\\
\indent Although early observations with limited spatial extent yielded radial density profiles consistent with an isothermal cylinder, the density profiles from recent observations with high dynamic range like those by the {\em Herschel Space Observatory}, especially of filaments located in nearby molecular clouds, are well described at large radii by considerably softer power laws than the isothermal case, i.e., $\alpha \approx 2$ or even smaller in Eq.\ (\ref{softpower}), indicating that these filaments are in a non-thermal state (\citealt{2015MNRAS.446.2110T}).

\subsection{Particle-based model}
\label{sec:particlemodel}

Our aim is to investigate how non-thermal stationary states may emerge from the dynamics. Hence, instead of studying the fluid equations (\ref{fluideqs}) and their stationary states, we use a particle-based model and study its dynamics by integrating its equations of motion. We still assume our system is infinitely extended and uniform along its longitudinal ($z$) axis, so that its dynamics is that of a system of infinite straight massive wires, whose mutual interaction is described by a two-dimensional (logarithmic) gravitational potential\footnote{We note that a similar model is used also in the context of unbunched charged particle beams in linear accelerators (see e.g.\ \citealt{1995PhRvE..51.3529R,1998PhRvS...1h4201W,2008PhRvL.100d0604L}, and references therein).} (see e.g.\ \citealt{1978MNRAS.184..709K,1994PhRvE..49.3771A,TelesLevinPakterRizzato:jstat2010}). Hence, the dynamics in a plane transverse to the $z$ direction is governed by the Hamiltonian of a two-dimensional system of $N$ self-gravitating particles,
\begin{equation}\label{2dgravhamiltonian}
\mathcal{H} = \sum_{i = 1}^N \frac{\left|\mathbf{p}_i\right|^2}{2m} + Gm^2 \sum_{i = 1}^N \sum_{j \not = i}^N \log \frac{\left|\mathbf{r}_i - \mathbf{r}_j \right|}{r_s}~,
\end{equation}
where $\mathbf{r}_i$ and $\mathbf{p}_i$ are, respectively, the position and momentum of the $i$-th particle in the plane, $r_s$ is an arbitrary scale length needed to make the argument of the logarithm dimensionless, and the mass $m$ of each particle is related to the total mass per unit length $M_\ell$ of the filament by $m = M_\ell/N$. The equations of motion to be integrated are thus
\begin{equation}\label{acc}
\ddot{\mathbf{r}}_i=-Gm\sum_{j\not = i}^N\frac{\mathbf{r}_i-\mathbf{r}_j}{\left|\mathbf{r}_i-\mathbf{r}_j\right|^2}~.
\end{equation}
Various aspects of the dynamics of such a system have been studied by \cite{2006CRPhy...7..331C}, \cite{TelesLevinPakterRizzato:jstat2010}, \cite{GabrielliJoyceMarcos:prl2010}, \cite{Marcos:pre2013}, \cite{LevinEtAlphysrep:2014}, \cite{MarcosGabrielliJoyce:pre2017}, \cite{2016PhLA..380..337S}.
For this model it is convenient to introduce dimensionless quantities in terms of scale radius $r_*$, dynamical time $t_*$ and scale velocity $v_*$.  We define the dynamical timescale of the system as
\begin{equation}
t_*\equiv\sqrt{\frac{2r_*^2}{GM_\ell}}\,,
\end{equation}
where $r_*=r_{50}(0)$ (i.e. the Lagrangian radius containing half of $M_\ell$ at $t=0$), and the velocity scale as
\begin{equation}
v_*\equiv r_{*}/t_*\,.
\end{equation}   

\subsection{Numerical approach}
\subsubsection{Direct $N$-body simulations}
\label{sec:nbody}

First of all, we performed direct $N$-body simulations where the acceleration on each particle was evaluated by direct sum. In this case, the divergence of the inter-particle potential and force for vanishing interparticle separation was regularized by introducing a softening length\footnote{The results we are going to present in the following do not depend on the choice of the softening length, at least as long as $10^{-5} \le \epsilon/r_* \le 10^{-2}$. For a theoretical study of the effect of a softening of power-law interactions on the existence and lifetime of nonequilibrium quasi-stationary states see \cite{GabrielliJoyceMarcos:prl2010} and \cite{MarcosGabrielliJoyce:pre2017}.} $\epsilon=10^{-3} r_*$, so that Eq.\ (\ref{acc}) actually reads
\begin{equation}\label{accsoft}
\ddot{\mathbf{r}}_i=-Gm\sum_{j\not = i}^N\frac{\mathbf{r}_i-\mathbf{r}_j}{\epsilon^2+\left|\mathbf{r}_i-\mathbf{r}_j\right|^2}~;
\end{equation}
the above equations were then integrated with a third-order symplectic algorithm  (see e.g. \citealt{1991CeMDA..50...59K,1995PhyS...51...29C}) with fixed timestep $\Delta t=t_*/100$ ensuring energy conservation up to 14 digits in double precision.\\
\indent In order to effectively explore different initial conditions, we limited ourselves to rather small $N$'s, that is, $N=3\times 10^4$. Our aim is to understand whether the purely gravitational dynamics of the system defined by the Hamiltonian (\ref{2dgravhamiltonian}) is able to produce non-thermal states similar to those observed in real filaments. Since the two-body relaxation time, after which ``collisional'' effects would eventually destroy any non-thermal steady state, is expected to grow\footnote{This happens in any long-range-interacting system; see e.g.\ \cite{CampaEtAl:book} for a general discussion and \cite{Marcos:pre2013}, \cite{TelesLevinPakterRizzato:jstat2010}, \cite{GabrielliJoyceMarcos:prl2010} and \cite{MarcosGabrielliJoyce:pre2017} for a discussion more focused on gravity and general power-law interactions.} with $N$, $N$ should be large enough to allow the formation and survival of such states. It has been shown by \cite{Marcos:pre2013} that the two-body relaxation time in a two-dimensional system of self-gravitating particles is of the order of $N t_*$, that is, a factor of $\ln N$ longer than in a three-dimensional system with the same $N$, while the time needed to form a non-thermal steady state should not depend on $N$ (and we shall see in Sec.\ \ref{sec:collapse} that such a timescale is not larger than $100t_*$). Hence, we expect that a number of particles of the order of $10^4$ is already sufficient to describe the physics we are interested in. This notwithstanding, it would be interesting to check whether our results do depend on $N$. In principle one should run $N$-body simulations with much larger $N$. Being such a task computationally demanding, to increase the total number of particles of three orders of magnitudes, up to $N = 1.5 \times 10^6$, we resorted to a standard two-dimensional particle-in-cell (PIC) method (to be described in Sec.\ \ref{sec:pic}) and to a novel method capable of effectively taking close encounters (collisions) into account in a PIC framework (Sec.\ \ref{sec:mpc}). These methods do not give a description of the dynamics of the model as accurate as the $N$-body method does, in that they do not properly model the physics at small length scales, but should still yield reasonable results, the more so because it has been shown by \cite{Marcos:pre2013} that in a two-dimensional self-gravitating system the slow relaxation processes that would eventually destroy the nonequilibrium stationary state are not dominated by the small-scale physics.
Moreover, comparing the results obtained by evolving the same initial conditions with different numerical methods may help us to gain some deeper insights in the results themselves (see Sec.\ \ref{sec:discussion}).

\subsubsection{Particle-in-cell simulations}
\label{sec:pic}
In a PIC approach, the system is represented by $N$ macro particles with mass $m_k$ moving in the plane. At every time step a fixed Cartesian grid of $N_c=N_x\times N_y$ cells is superimposed to the system, so that the mesh based mass density is obtained as
\begin{equation}\label{rhocont}
\rho_{i,j}=\tilde\rho_{i,j}+\frac{1}{\Delta x\Delta y}\sum_{k=1}^{n_{i,j}}m_k\,,
\end{equation} 
where $\Delta x$ and $\Delta y$ are the cell's side lengths, $n_{i,j}$ is the number of particles enclosed by the cell, and $\tilde\rho$ is a static background density. In our implementation the particles' contribution to the density in a cell is evaluated with the simple nearest-grid-point method (NGP, \citealt{1981csup.book.....H}).
In order to compute the resulting gravitational potential, the code solves the Fourier-space based Poisson equation
\begin{equation}
\hat\Phi_{i,j}=\hat\rho_{i,j}\hat{\mathcal{G}}_{i,j}\,,
\end{equation}
where the hats over the quantities denote the Fourier transforms, and $\mathcal{G}$ is the Green's function of the Laplace operator. Once the mesh based potential $\Phi_{i,j}$ is recovered by antitransforming $\hat\Phi_{i,j}$, the gravitational acceleration $\mathbf{a}(\mathbf{r}_k)=-\nabla\Phi(\mathbf{r}_k)$ acting on a particle sitting at position $\mathbf{r}_k$ inside the cell of indices $i,j$, is obtained via the gradient interpolation (see \citealt{2000NewA....5..305F})
\begin{equation}
\begin{split}
\frac{\partial\Phi}{\partial x}(\mathbf{r}_k)=\frac{\Phi_{i+1,j}-\Phi_{i-1,j}}{2\Delta x}+\frac{\Phi_{i+1,j}+\Phi_{i-1,j}-2\Phi_{i,j}}{\Delta x^2}\delta x_k\\
+\frac{\Phi_{i+1,j+1}-\Phi_{i-1,j+1}+\Phi_{i-1,j-1}-\Phi_{i+1,j-1}}{4\Delta x\Delta y}\delta y_k\,.
\end{split}
\end{equation}
In the equation above (and in the analogous one for the $y$ component of $\nabla\Phi$), $\delta x_k$ and $\delta y_k$ are the components of $\mathbf{r}_k-\mathbf{c}_{i,j}$, where $\mathbf{c}_{i,j}$ is the position of the cell's center.
As in the $N$-body case, the equations of motion of the particles are integrated with a third order symplectic integrator scheme with fixed timestep $\Delta t=t_*/100$ ensuring energy conservation up to 14 digits in double precision. For the simulations reported here we have used $N=1.5\times 10^6$ particles on a Cartesian grid with $N_c=128\times 128$ equal square cells.

\subsubsection{Inclusion of collisions}
\label{sec:mpc}
One of the limits of a PIC approach is that it underestimates the contribution of close encounters (collisions) between particles, that are included in the computationally heavier $N$-body simulations, at least up to a distance scale of order $\epsilon$. To effectively overcome this limitation, following a suggestion of \cite{2010JPhCS.260a2005B,2013PhRvE..87b3102B} and \cite{2015PhRvE..92f2108D,2017PhRvE..95d3203D} in the context of numerical modelling of plasma transport, we treat collisional processes (i.e.\  particle scatterings) with the stochastic multi-particle collision scheme (hereafter MPC), originally introduced by \cite{1999JChPh.110.8605M} in the field of mesoscopic fluid dynamics (see also \citealt{2004LNP...640..116M,kapral08,2009acsa.book....1G}).
The MPC methods simulate inter-particle collisions at the cell level by performing a stochastic rotation of their velocity vectors $\mathbf{v}_k$, constrained by the conservation of the total momentum and kinetic energy of the cell. In a two dimensional implementation, at the beginning of each timestep $\delta t$, in the cell of indexes $i,j$ first of all the center of mass velocity is evaluated as
\begin{equation}\label{mtotptot}
\mathbf{u}_{ij}=\frac{1}{n_{ij}}\sum_{k=1}^{n_{ij}}\mathbf{v}_k\,.
\end{equation}  
Then an angle $\vartheta_{ij}$ is sampled from a uniform distribution in $(0,2\pi)$. The collision itself is simulated by rotating with probability one-half the relative velocities $ \delta\mathbf{v}_k=\mathbf{v}_k-\mathbf{u}_{ij}$, as
\begin{equation}\label{rotation}
\mathbf{v}_{k}^\prime=\mathbf{u}_{ij}+\mathcal{R}_{ij}\cdot\delta\mathbf{v}_{k}\,,
\end{equation}  
where $\mathcal{R}_{ij}$ is the 2D rotation matrix of an angle $\vartheta_{ij}$. 
Such a rotation guarantees the conservation of the total momentum
\begin{equation}\label{sist}
\mathbf{P}_{ij}=\sum_{k=1}^{n_{ij}} m\mathbf{v}_k = \sum_{k=1}^{n_{ij}} m\mathbf{v}_k^\prime
\end{equation}
and kinetic energy
\begin{equation}\label{sist1}
K_{ij}=\frac{1}{2}\sum_{k=1}^{n_{ij}} m\mathbf{v}_k^2=\frac{1}{2}\sum_{k=1}^{n_{ij}} m\mathbf{v}_k^{\prime 2}
\end{equation}
in the cell. By doing so, however, the total $z$-component $L_z$ of the angular momentum  of the cell is not conserved. To overcome this inconvenience, following \cite{tesiryder} in our implementation of the MPC rule, we introduce the following constraints on $\vartheta_{ij}$ 
\begin{equation}\label{sincosphi}
\sin\vartheta_{ij}=-\frac{2a_{ij}b_{ij}}{a_{ij}^2+b_{ij}^2}\,;\quad  \cos\vartheta_{ij}=\frac{a_{ij}^2-b_{ij}^2}{a_{ij}^2+b_{ij}^2}\,.
\end{equation}
The latter hold if the coefficients $a_{ij}$ and $b_{ij}$ are given as functions of particles' positions and velocities in the cell by
\begin{equation}\label{adef}
a_{ij}=\sum_{k=1}^{n_{ij}}\mathbf{r}_k \wedge (\mathbf{v}_k-\mathbf{u}_{ij})\,;\quad
 b_{ij}=\sum_{k=1}^{n_{ij}}\mathbf{r}_k\cdot(\mathbf{v}_k-\mathbf{u}_{ij})\,.
\end{equation}
Since the effective collisionality may in principle depend on the local state of the system (i.e., particles' velocities and density), the collision move is accepted only if a random probability $\mathcal{P}^*_{ij}$ sampled from a uniform distribution in $(0,1)$ is smaller than the cell dependent collision probability
\begin{equation}
\mathcal{P}_{ij}=1-\exp\left[-(v_{ij}\Delta t n_{ij}d_{ij}/\Delta x\Delta y)^2\right]\,.
\end{equation}
In the expression above, $v_{ij}$ and $d_{ij}$ are the average velocity and average interparticle distance in the cell, respectively. Note that the collision probability has no adjustable parameters. It must be noted that the ``collisions'' taken into account by this PIC+MPC scheme are not the ``true'' ones dictated by the gravitational interactions, but only a stochastic effective model of the latter.

\section{Cold collapses}
\label{sec:collapse}
\begin{figure}
\centerline{\includegraphics[width=0.9\columnwidth]{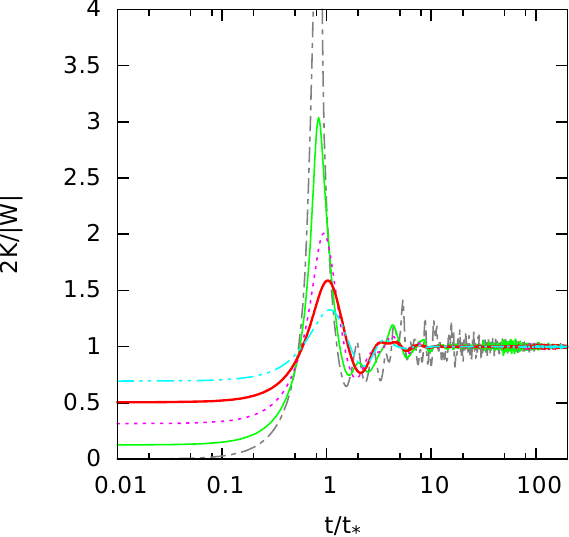}}
\caption{Time evolution of the virial ratio $2K/|W|$ for systems starting with Gaussian density profiles with $r_0 = 1$ and the following initial values of $2K/|W|$: 0.0 (grey dot-dashed line), 0.1 (green thin solid line), 0.3 (red dotted line), 0.5 (red thick solid line) and 0.7 (cyan dot-dot-dashed line). MPC+PIC simulations with $N = 1.5\times 10^6$.}
\label{fig_virial}
\end{figure}

\begin{figure*}
\centerline{\includegraphics[width=0.95\textwidth]{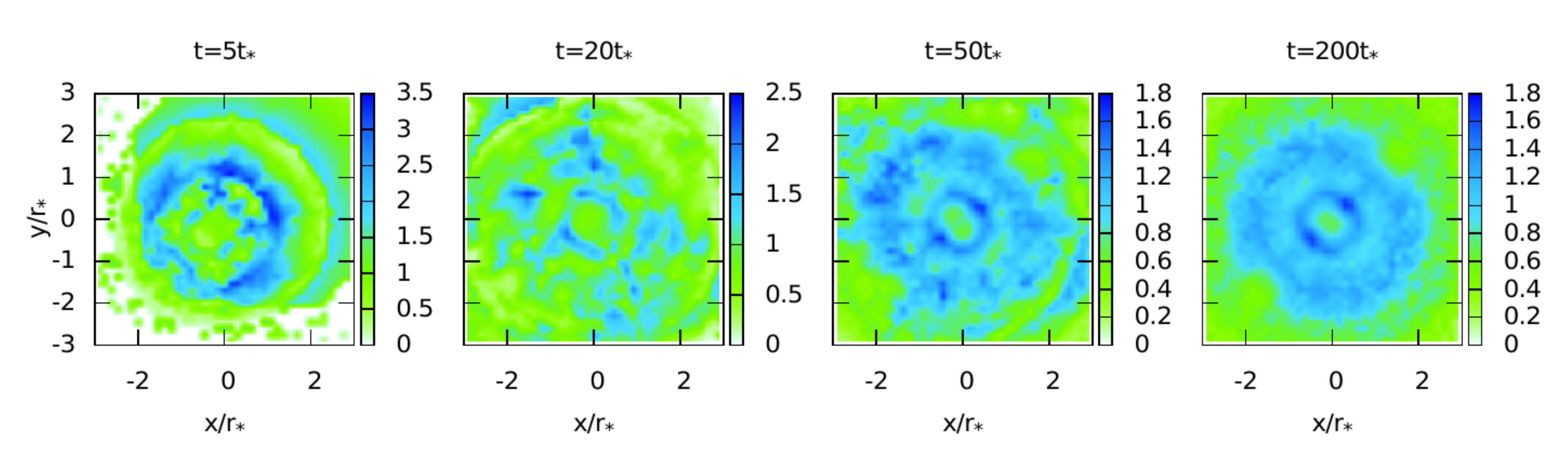}}
\caption{Local kinetic temperature (colour scale) as a function of the position of an initially perfectly cold (with vanishing initial virial ratio, corresponding to the dot-dashed curve in Fig.\ \ref{fig_virial}) Gaussian filament with $r_0 = 1$ at, from left to right, $t=5t_*$, $20t_*$, $50t_*$ and $200t_*$. The normalization of lengths is given in units of the initial half mass radius $r_*$. PIC-MPC simulations with $N = 1.5\times 10^6$.}
\label{fig_Tevol}
\end{figure*}

The first scenario we investigate for the dynamical generation of non-thermal states in our simple model of a filament is that of cold collapse. As initial condition for the positions we assume a cylindrically symmetric Gaussian overdensity with initial radial density profile given by
\begin{equation}\label{rhoexp}
\rho(r)=\rho_c\exp\left(-\frac{r^2}{2r_0^2}\right)\,,
\end{equation}
where the central density $\rho_c$ is taken to be ten times larger than the homogeneous background $\rho$.  The initial particle velocities are extracted from a position independent Maxwellian such that the initial virial ratio of the system is $(2K/|W|)_{t=0}<1$, where 
\begin{equation}\label{cinetica}
K=\sum_{i=1}^{N}\frac{m \left|\mathbf{v}_{i}\right|^{2}}{2}
\end{equation}
is the total kinetic energy of the system, and 
\begin{equation}\label{viriale1}
W = \sum_{i = 1}^N \mathbf{r}_i \cdot \mathbf{F}_{i}
\end{equation}
is the virial function, where 
\begin{equation}\label{F}
\mathbf{F}_i=-Gm^2\sum_{j\not = i}^N\frac{\mathbf{r}_i-\mathbf{r}_j}{\left|\mathbf{r}_i-\mathbf{r}_j\right|^2}
\end{equation}
is the force exerted on the $i$-th particle. We note that in the two-dimensional case $W$ does not depend on time,  
\begin{equation}\label{eq:W}
W = -\frac{Gm^2}{2}N(N-1)\,.
\end{equation}
In the following we report on the results obtained with the Gaussian initial profiles (\ref{rhoexp}), but those with the truncated initial distribution look almost the same. \\
\indent As expected, at small times the system undergoes a phase of violent relaxation with strong oscillations of the virial ratio, stronger when the initial value of $2K/|W|$ is smaller (Fig.\ \ref{fig_virial}). Virial oscillations damp out on a timescale of some tens of dynamical times $t_*$, and at times $t > 100 t_*$ the system appears to be in an essentially steady state. In Fig.\ \ref{fig_virial} only the results obtained with the MPC+PIC simulation protocol are reported but those obtained with simple PIC without collisions and $N$-body simulations are very similar.
Monitoring the kinetic temperature as a function of the position (i.e., the locally averaged kinetic energy) during the time evolution for a completely cold (i.e., with vanishing initial virial ratio) collapse, we observe that it wildly changes exhibiting highly irregular patterns during the violent relaxation phase, but then settles in an azimuthally symmetric pattern where temperature inversion is clearly evident, as shown in Fig.\ \ref{fig_Tevol}.
\subsection{Density and temperature profiles}
\label{sec:profiles}
\begin{figure*}
\centerline{\includegraphics[width=0.9\textwidth]{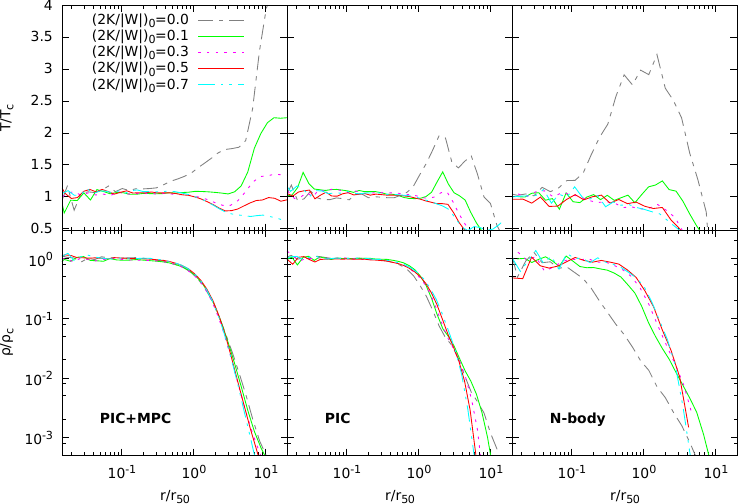}}
\caption{Azimuthally averaged kinetic temperature (top) and density (bottom) radial profiles at $t=200t_*$ for systems with initial Gaussian density profile, $r_0=1$, and different values of the initial virial ratio (see legend). From left to right the panels shows the results of simulations performed with PIC+MPC method, simple PIC and, finally, direct $N$-body. In the first two cases $N=1.5\times 10^{6}$ particles have been used, while in the last $N=3\times10^4$. The radii are expressed in units of the Lagrangian radius $r_{50}$ at $t=200t_*$ while temperatures and densities are given in units of their central values $T_c$ and $\rho_c$, respectively.}
\label{fig_tempcoll}
\end{figure*}

\begin{figure}
\centerline{\includegraphics[width=0.85\columnwidth]{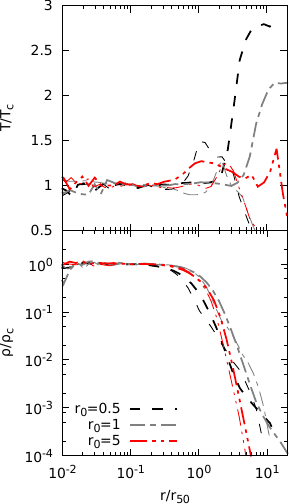}}
\caption{Azimuthally averaged kinetic temperature (top) and density (bottom) radial profiles at $t=200t_*$ for systems starting with Gaussian density profile with $(2K/|W|)_0=0.1$ and different values of $r_0$ (see legend), with collisions included (thick lines) and not included (thin lines). In all cases $N=1.5\times 10^6$.}
\label{fig_collisionnocollision}
\end{figure}

\begin{figure*}
\centerline{\includegraphics[width=0.95\textwidth]{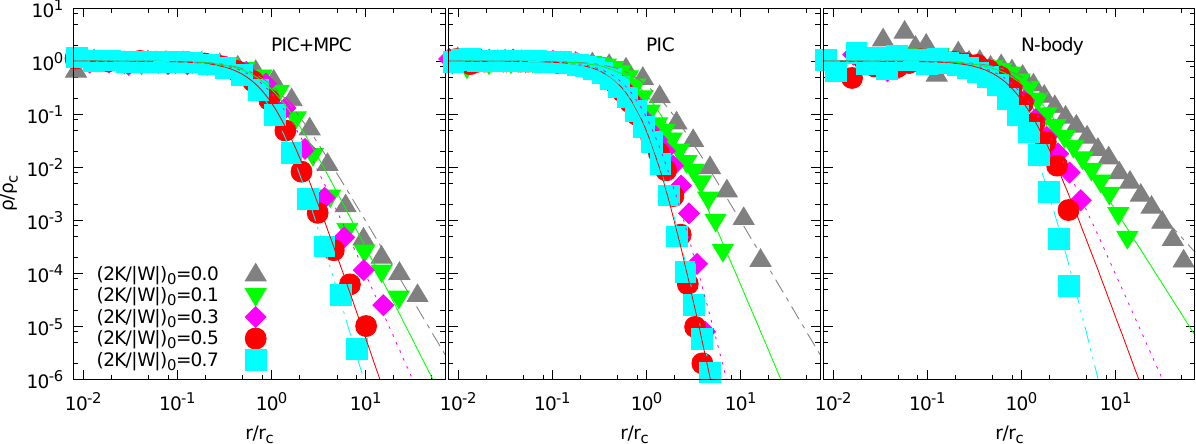}}
\caption{Azimuthally averaged density profiles at $t=200t_*$ after the collapse of initially Gaussian states with $r_0 = 1$ and with different initial virial ratios (symbols, see legend) and best-fit curves according to Eq.\ (\ref{softpower}) (lines). Left panel: MPC+PIC simulations with $N = 1.5 \times 10^6$, curves with $\alpha=3.0$, 3.5, 4.0, 5.2 and 6.1 from right to left. Center panel: simple PIC simulations with $N = 1.5 \times 10^6$, curves with $\alpha=3.0$, 4.2, 7.0, 8.7 and 8.5 from right to left. 
Right panel: $N$-body simulations with $N = 3 \times 10^4$, curves with $\alpha=1.9$, 2.8, 4.0, 4.8 and 7.3 from right to left.
Densities are given in units of the numerically evaluated core density $\rho_c$ and radii in terms of the corresponding core radius $r_c$ as given by Eq.\ (\ref{softpower}).}
\label{fig_fit1}
\end{figure*}

The density and kinetic temperature (azimuthally averaged) radial profiles in the steady states reached after a cold collapse of a Gaussian initial profile with $r_0 = 1$ are shown in Fig.\ \ref{fig_tempcoll}, for all the three types of simulations we performed, PIC+MPC, PIC without collisions, and direct summation $N$-body. The qualitative properties of the density profiles are very similar: all exhibit a flat, essentially isothermal core and decay as a power law at larger radii. The power law is steeper for initially warmer and softer for initially colder systems, that also exhibit temperature inversion, while initially warmer systems do not show this feature and rather their kinetic temperature decreases with radius. Temperature inversion does depend on the simulation method: it is stronger in the $N$-body and especially in the PIC+MPC cases, while it is much less evident in the PIC without collisions simulations. 
The properties of the final steady state also depend on the width of the initial Gaussian profile, as shown in Fig.\ \ref{fig_collisionnocollision} for MPC+PIC and simple PIC simulations: as a general rule, regardless of the presence of collisions, initially more concentrated systems produce softer decays at large radii and stronger temperature inversions in the final state. 
Remarkably, as shown in Fig.\ \ref{fig_fit1}, the density profiles in the final steady state are quite well described by the empiric relation\footnote{This may suggest that these states are indeed essentially polytropic, but an analysis of the velocity anisotropies (see Sec.\ \ref{sec:anisotropy}) shows that this may be an oversimplified description.} (\ref{softpower}). The exponent $\alpha$ depends on the initial virial ratio and encompasses a broad range of values that includes the thermal value $\alpha = 4$, being smaller for initially colder systems and varying from $\alpha = 1.9$ for a completely cold collapse in an $N$-body simulations to $\alpha = 8.7$ for a collapse with initial virial ratio equal to 0.5 in a PIC simulation. 
The values of the fitted $\alpha$ exponent are also reported in Table \ref{table:alpha} in the various cases of cold collapses of Gaussian initial states with $r_0 = 1$.
\begin{table}
\begin{tabular}{c|ccc}
$2K/|W|$ (t = 0) & $\alpha$ (MPC+PIC) & $\alpha$ (PIC) & $\alpha$ ($N$-body) \\
\hline \hline
0.0 & 3.0 & 3.0 & 1.9 \\
0.1 & 3.5 & 4.2 & 2.8 \\
0.3 & 4.0 & 7.0 & 4.0 \\
0.5 & 5.2 & 8.7 & 4.8 \\
0.7 & 6.1 & 8.5 & 7.3 \\
\end{tabular}
\caption{Values of the $\alpha$ exponent obtained fitting Eq.\ (\ref{softpower}) to the final radial density profiles for cold collapses of initially Gaussian profiles with $r_0 = 1$.} 
\label{table:alpha}
\end{table}
\subsection{Velocity distribution anisotropy}
\label{sec:anisotropy}
\begin{figure*}
\centerline{\includegraphics[width=0.9\textwidth]{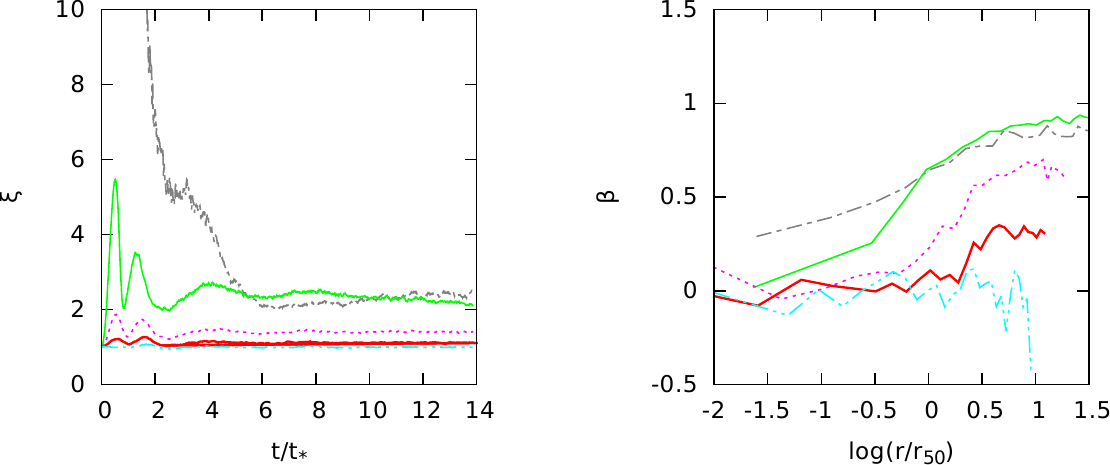}}
\caption{(Left panel) Time evolution of the global velocity anisotropy parameter $\xi$ defined in Eq.\ (\ref{eq:xi}). (Right panel) Anisotropy profile $\beta$, defined in Eq.\ (\ref{eq:beta}), as a function of the rescaled radius $r/r_{50}$ in the non-thermal end state at $t = 200t_*$. All the curves have been obtained via direct $N$-body simulations of collapses of initially Gaussian overdensities with $r_0 = 1$ and $N = 3\times 10^4$. In both panels, the different curves correspond to the following initial values of $2K/|W|$: 0.0 (grey dot-dashed line), 0.1 (green thin solid line), 0.3 (red dotted line), 0.5 (red thick solid line) and 0.7 (cyan dot-dot-dashed line), that is, the same color code as in Figs.\ \ref{fig_virial} and \ref{fig_tempcoll}.}
\label{fig_anisotropy}
\end{figure*}

Further insight into the properties of the violent relaxation process at the beginning of the collapse as well as into the features of the non-thermal state reached after the damping of the virial oscillations can be obtained by studying the anisotropies of the velocity distribution. Were the system in a thermal equilibrium state, the velocity distribution would be isotropic, but also non-thermal states can be isotropic, e.g., when the single-particle distribution function only depends on the total energy. To this end, in our direct $N$-body simulations we adopted standard diagnostic tools to reveal possible anisotropies. First, we measured the time evolution of the global anisotropy parameter (see e.g.\ \citealt{TrentiBertin:Apj2006} and references therein)
\begin{equation}
\xi = \frac{K_r}{K_\varphi}\,,
\label{eq:xi}
\end{equation}
where $K_r$ and $K_\varphi$ are the radial and tangential kinetic energies, respectively. In a system where the orbits are mostly radial (resp.\ tangential), $\xi$ is large (resp.\ small); an isotropic velocity distribution yields $\xi = 1$. After the system had reached its steady state, we measured the azimuthally averaged radial profile of the anisotropy parameter  
\begin{equation}
\beta(r) = 1 - \frac{\sigma^2_\varphi(r)}{\sigma^2_r(r)}~,
\label{eq:beta}
\end{equation}
where $\sigma_r$ and $\sigma_\varphi$ are the radial and tangential velocity dispersions, respectively. The parameter\footnote{Not to be confused with the $\beta$ entering Eq.\ \ref{rcostriker}, that is related to the temperature of the isothermal cylinder.} $\beta$ measures the pressure anisotropy as a function of the radius $r$, as well as the kinematical structure of the particles' orbits \citep{2008gady.book.....B}: in an isotropic state $\beta = 0$, while positive (resp.\ negative) values of $\beta$ denote a radial (resp.\ tangential) anisotropy. Purely radial orbits yield $\beta = 1$, while perfectly circular orbits would give $\beta = -\infty$. 
The results are shown in Fig.\ \ref{fig_anisotropy}. 
The time evolution of $\xi$ shows that the global velocity anisotropy is rather large at the beginning of the collapse (colder collapses are more anisotropic) and decreases with the damping of the virial oscillations; in warmer collapes the global amount of anisotropy is negligible at large times, while it remains quite relevant for initially colder states. The $\beta(r)$ profiles in the non-thermal steady state are all monotonously increasing with radius (with the only exception of the warmer collapse) and show that in the outer regions anisotropy is always present and low-angular momentum orbits dominate (again with the exception of the warmer collapse where the anisotropy becomes tangential at large radii).
The fact that initially warmer states are also more isotropic, both in their time evolution and in their end states, is expected due to the larger amount of random motions with respect to the initially colder cases. The rather large amount of radial anisotropy found in the initially colder cases, especially in the outer regions, may be due to the fact that in cold collapses the velocities are initially extremely anisotropic, because in the initial stages of virial oscillations the motion is predominantly radial when starting from cylindrically symmetric initial conditions (this is witnessed by the fact that $\xi$ is very large at small times)\footnote{We note that in two dimensions we may expect larger anisotropies than in three dimensions, due to the smaller number of available tangential channels for the velocity to relax to an isotropic distribution.}. Anisotropy is thus generated at the beginning of the virial oscillations and decreases during the relaxation to the non-thermal end state, but only in warmer collapses this process efficiently reduces the initial anisotropy to a small amount\footnote{In three-dimensional collapses it has been shown that starting from highly symmetric initial conditions results in more anisotropic end states than in the case of less symmetric initial states \citep{TrentiBertin:Apj2006,vanAlbada:mnras1982}.}. Since the initially colder collapses are also those where temperature inversion is more pronounced, we observe a correlation between temperature inversion and velocity anisotropy in our simulations; however, our results alone are not sufficient to suggest that a correlation between temperature inversion and anisotropy should hold in general in a low-dimensional self-gravitating system. \\
\indent As mentioned in Sec.\ \ref{sec:fluidmodels}, density profiles given by Eq.\ (\ref{softpower}) are very close to those obtained in a fluid picture assuming a polytropic equation of state. However, the latter states are isotropic, hence our results on velocity anisotropy suggest that an interpretation of the non-thermal states of filaments in terms of purely polytropic states may be an oversimplified picture, although it correctly describes the radial density profile.     

\section{Perturbation of a thermal state}
\label{sec:perturb}
\begin{figure}
\centerline{\includegraphics[width=0.9\columnwidth]{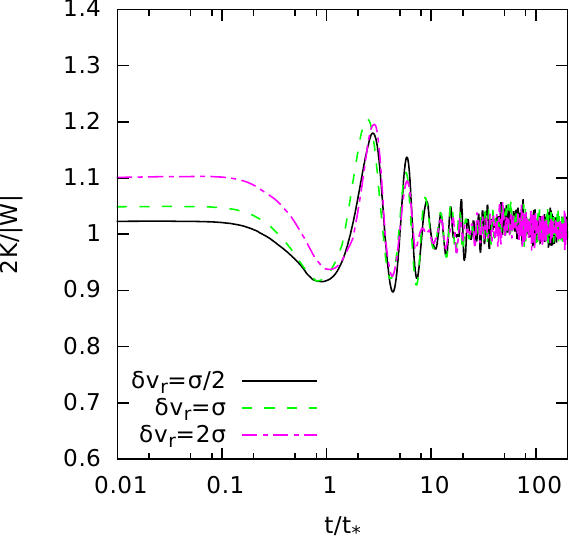}}
\caption{Time evolution of the virial ratio $2K/|W|$ for systems starting with initial conditions corresponding to an Ostriker equilibrium model with applied radial perturbations of different strength (lines), obtained with MPC+PIC simulations with $N = 1.5 \times 10^6$; $\sigma$ is the velocity dispersion of the initial state.}
\label{fig_virialpert}
\end{figure}

\begin{figure}
\includegraphics[width=0.85\columnwidth]{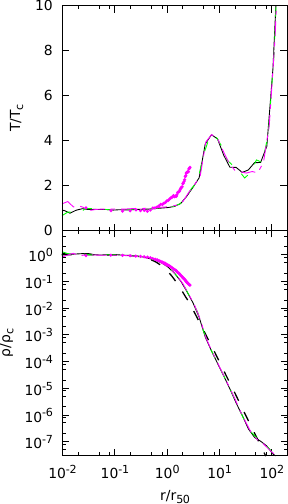}
\caption{Azimuthally averaged kinetic temperature (top) and density (bottom) profiles at $t=200t_*$ for Ostriker filaments undergoing a radial perturbations. Lines refer to the same cases as in Fig.\ \ref{fig_virialpert}, while symbols refer to an $N$-body simulation with $N = 10^4$ and $\delta v_r = 2\sigma$. Densities and temperatures are normalized to their central values and radii to $r_{50}$. The thick dashed line in the bottom panel marks the initial density profile given by Eq.\ (\ref{ostriker}).
}
\label{fig_ostriker}
\end{figure}

The second dynamical scenario we consider is the evolution of a strong impulsive radial perturbation of an initially isothermal profile, i.e., given by the Ostriker-Stod\'olkiewicz self-gravitating cylinder--Eq.\ (\ref{softpower}) with $\alpha = 4$. Physically, this scenario may model, albeit in a very simplified way, the evolution of an already thermally relaxed filament that is suddenly perturbed by a shock wave. Moreover, such scenario is interesting because it bridges our results to other recent results obtained for models of long-range-interacting systems in non-astrophysical contexts (\citealt{2014EPJB...87...91C,2015PhRvE..92b0101T,2016PhRvE..93f6102T,njp2016}), where perturbations of initially thermal states were considered.\\
\indent The density profile of the isothermal cylinder is given by setting $\alpha = 4$ in Eq.\ (\ref{softpower}), thus obtaining 
\begin{equation}\label{ostriker}
\rho(r)=\frac{r_c^4\rho_c}{(r_c^2+r^2)^2}\,,
\end{equation}
so that the associated radial mass per unit length profile becomes
\begin{equation}\label{massprofost}
M_\ell(r)=\frac{\pi \rho_c r^2r_c^2}{r^2+r_c^2}=\frac{M_\ell r^2}{r^2+r_c^2}\,,
\end{equation}
where the core radius $r_c$ is linked to $T$ through Eq.\ (\ref{rcostriker}) and $r_*=r_c$, see also \cite{1964ApJ...140.1056O}. Initial particle positions are obtained by sampling Eq.\ (\ref{massprofost}) for the radial coordinate $r$ and then assigning an angular coordinate $\varphi$ form a uniform distribution in the interval $(0,2\pi)$. After this, the velocity components $v_r$ and $v_\varphi$ are extracted consistently from the position independent isotropic Maxwellian at temperature $T$
\begin{equation}\label{maxwell}
f(v)=\sqrt{\left(\frac{m}{2\pi k_B T}\right)^3}4\pi v^2e^{-\frac{mv^2}{2k_BT}};\quad v=\sqrt{v_r^2+v_\varphi^2},
\end{equation}
and finally converted in Cartesian coordinates along with their positions components.
In order to simulate a radial compression, the radial component of particles' velocities  $v_r$ is instantaneously increased of a quantity $\delta v_r\approx 2\sqrt{k_B T}$. Alternatively, in some simulations particles' positions have been displaced of a quantity $\delta_r\approx r_*/80$ along their radial coordinate; the results are very similar to those obtained with the velocity displacement, and in the following we report only on the latter.\\
\indent The time evolution of the virial ratio for different strengths of the perturbation is shown in Fig.\ \ref{fig_virialpert}. As in the cold collapses, the system undergoes virial oscillations, that damp out on a timescale of tens of dynamical times, and then settles in a steady state. Kinetic temperature and density profiles of such a state are shown\footnote{Preliminary results limited to the $N$-body case were already presented by \cite{2015PhRvE..92b0101T}.
} in Fig.\ \ref{fig_ostriker}. It is apparent that the kinetic temperature and density radial profiles are anticorrelated: temperature inversion is even stronger than in the completely cold collapse case. As far as the relation between temperature and density is concerned, the steady states reached after the perturbation of a thermal cylinder appear as polytropic, at least when density is not too small: indeed, the $\rho$ versus $T$ relation is reasonably fitted by a power law (data not shown), 
and in general the quality of the fit improves for systems undergoing stronger perturbations. However, also in this case there are velocity anisotropies (data not shown), so that these states can not be considered as ``true'' polytropic states. 

\section{Discussion}
\label{sec:discussion}
As shown in Sec.\ \ref{sec:profiles}, the steady states reached by our simple model of a filament after a cold collapse are definitely non-thermal, unless the initial condition is fine-tuned, and qualitatively very similar to the observed states, as their density profiles are in good agreement with Eq.\ (\ref{softpower}) with values of the exponent $\alpha$ that depend on the initial condition, being in general smaller for colder collapses. Another indication of the fact that the steady states are non-thermal is the observed velocity anisotropy (Sec.\ \ref{sec:anisotropy}). Moreover, colder collapses exhibit appreciable temperature inversion. The fact that such a simple model of a filament, neglecting all the dissipative effects, rotation, radiation, and magnetic fields, and retaining only self-gravity, exhibits steady states that are so similar to the observed ones suggests that the dynamics alone is sufficient to produce and support the non-thermal states observed in real filaments, and that other mechanisms are not strictly necessary. Moreover, the fact that all the three numerical methods ($N$-body, PIC and PIC+MPC) yield qualitatively similar results is an indication of the robustness of the results themselves.  Quantitatively, the values of the exponent $\alpha$ governing the decay of the radial density profile at large radii obtained for colder collapses with $N$-body simulations seem compatible with those observed in real filaments (\citealt{2015MNRAS.446.2110T}), while all the other cases yield values that are larger than the observed ones (see Table \ref{table:alpha}). Still, the values of $\alpha$ depend not only on the initial conditions, but also on the choice of the simulation protocol. First of all, we note that the values of $\alpha$ closer to the observed ones are obtained by means of $N$-body simulations. This is an important result because the $N$-body method is the most correct one  among the three considered, since it captures the physics at all length scales (larger than the softening length). Then, as witnessed by the values reported in Table \ref{table:alpha}, we observe that the general trend, i.e., colder collapses corresponding to final profiles decaying with softer power laws, is the same for all the three simulation techniques, but $N$-body and PIC simulations have the same spread in the values of $\alpha$ while the $\alpha$'s obtained with MPC+PIC simulations are closer to each other; moreover, as far as absolute values are concerned, $N$-body simulations yield the smallest $\alpha$'s, at least for colder collapses, while the values found in PIC simulations are the largest. This may suggest that the small-scale physics, including close encounters between particles, that is somewhat suppressed in PIC simulations and (only approximately, via a stochastic model) restored in MPC+PIC ones, is important for the processes that produce ``hot'' particles able to create a more diffuse halo at large radii. Consistently, temperature inversion is stronger in states obtained with MPC+PIC and $N$-body simulations, while is suppressed in PIC simulations. On the other hand, the fact that the values of $\alpha$ obtained with MPC+PIC simulations span a smaller interval and are closer to the thermal value may be a consequence of that including random collisions allows the system to get closer to thermal than in the other cases.\\  
\indent As recalled in the Introduction, self-gravitating systems do not reach thermal equilibrium (that in the case of two-dimensional systems exists also for a finite system, at variance with the three-dimensional case) when starting from nonequilibrium initial conditions because the time need to reach a thermal state is exceedingly long and grows with the number of particles $N$. The steady states we observe after the violent relaxation phase are rather non-thermal, quasi-stationary states that would become exactly stationary in the limit $N\to\infty$. Theoretically, such states are identified with stable stationary solutions of the Vlasov (or collisionless Boltzmann) equation, that describes the evolution of the single-particle distribution function for such systems for time scales smaller than the relaxation time (that becomes infinite in the $N\to\infty$ limit). In the last decades, it has been recognized that the presence of such quasi-stationary, non thermal states as well as of long relaxation times growing with $N$ and preventing an effective relaxation to thermodynamic equilibrium are common to all many-body systems with long-range interactions and not restricted at all to self-gravitating systems (\citealt{CampaEtAl:book}; see also \citealt{2013MNRAS.431.3177D,2011IJBC...21.2279D}). It has been recently suggested that non-thermal states with temperature inversion are a typical outcome when a generic many-body long-range-interacting system relaxes to a quasi-stationary states after undergoing collective oscillations (\citealt{2014EPJB...87...91C,2015PhRvE..92b0101T,2016PhRvE..93f6102T,njp2016}). A possibly ``universal'' mechanism has been proposed to explain the ubiquity of temperature inversion by \citet{2015PhRvE..92b0101T}. The mechanism can be summarized as follows: during the initial violent relaxation phase the interaction of the particles with the collective oscillations, that will eventually lead to the Landau-like damping of the oscillations themselves, may produce suprathermal tails in the velocity distribution function even when starting from a thermal state. In an inhomogeneous system, this may trigger a ``velocity filtration'' process (\citealt{1992ApJ...398..299S,1992ApJ...398..319S}) that effectively broadens the velocity distribution function in the regions where the system is less dense, because only sufficiently fast particles can escape the potential well produced by the central concentration. \\
\indent The systems studied by \cite{2015PhRvE..92b0101T} and \cite{njp2016} were initially in a thermal state, and then brought out of equilibrium by a sudden perturbation or a quench of an external parameter. The results we have shown in Sec.\ \ref{sec:perturb} confirm the general picture also for two-dimensional gravitational systems. The fact that also cold collapses, where the starting configuration is not a thermal one, may exhibit temperature inversion is a significant addition to this picture: it indeed suggests that collective oscillations and wave-particle interactions are the crucial ingredient to produce a final state with temperature inversion, although not a sufficient one. Our results with MPC+PIC simulations then suggest that collisions may even enhance the temperature inversion.

\section{Concluding remarks}
\label{sec:conclusions}
The main result of the work we have presented here is that a very simple model of a filament, that completely neglects dissipative effects, radiation, magnetic fields, and rotation, retaining only gravity and effectively mapping the dynamics of a filament onto that of a two-dimensional system of self-gravitating particles, is able to produce non-thermal steady states similar to those observed in real filaments, as end states emerging after the damping of virial oscillations in cold collapses. The radial density profiles of such states are well described by Eq.\ (\ref{softpower}) and for sufficiently cold collapses $N$-body simulations yield values of the exponent $\alpha$ governing the decay of the density profile at large radii in the range of the observed ones. \\
\indent Clearly, our results do not imply that mechanisms other than dissipationless dynamics alone are not at work and that they are not important in the real case; all the effects and processes we neglected in our very simple models may be important and a strategy to understand their r\^{o}le and relevance may be the progressive addition of other ingredients to our simple model, to see whether they may bring the non-thermal final states closer or not to the observed ones. The first step is naturally the inclusion of magnetic fields, that couple to the small fraction of ionized molecules in the filament and whose effect is thus transferred to the whole system by the interactions between the ionized particles and all the others. Work is in progress in this direction. In the fluid picture, polytropic solutions for infinite cylinders with magnetic fields were analyzed by \cite{2015MNRAS.446.2118T}, while the effect of rotation and of an imposed temperature gradient were considered by \cite{2013A&A...558A..27R,2014MNRAS.444.1775R}. \\
\indent Finally, the non-thermal states observed as end products of the dynamics of our simple model exhibit temperature inversion, that seems a common feature of a wide class of non-thermal states of any long-range-interacting system, hinting at a possible and intriguing relation between non-thermal states of galactic filaments and properties of nonequilibrium states that could be observed in other systems, even in condensed matter ones.

\section*{Acknowledgments}
We thank Guido Ciraolo, Daniele Galli, and Tarc\'{\i}sio N.\ Teles for insightful discussions and important comments at an early stage of this work. We also thank an anonymous referee for very useful comments that helped us to improve the presentation of our work.
\bibliographystyle{mnras}
\bibliography{biblio.bib}

\begin{thebibliography}{}
\makeatletter
\relax
\def\mn@urlcharsother{\let\do\@makeother \do\$\do\&\do\#\do\^\do\_\do\%\do\~}
\def\mn@doi{\begingroup\mn@urlcharsother \@ifnextchar [ {\mn@doi@}
  {\mn@doi@[]}}
\def\mn@doi@[#1]#2{\def\@tempa{#1}\ifx\@tempa\@empty \href
  {http://dx.doi.org/#2} {doi:#2}\else \href {http://dx.doi.org/#2} {#1}\fi
  \endgroup}
\def\mn@eprint#1#2{\mn@eprint@#1:#2::\@nil}
\def\mn@eprint@arXiv#1{\href {http://arxiv.org/abs/#1} {{\tt arXiv:#1}}}
\def\mn@eprint@dblp#1{\href {http://dblp.uni-trier.de/rec/bibtex/#1.xml}
  {dblp:#1}}
\def\mn@eprint@#1:#2:#3:#4\@nil{\def\@tempa {#1}\def\@tempb {#2}\def\@tempc
  {#3}\ifx \@tempc \@empty \let \@tempc \@tempb \let \@tempb \@tempa \fi \ifx
  \@tempb \@empty \def\@tempb {arXiv}\fi \@ifundefined
  {mn@eprint@\@tempb}{\@tempb:\@tempc}{\expandafter \expandafter \csname
  mn@eprint@\@tempb\endcsname \expandafter{\@tempc}}}

\bibitem[\protect\citeauthoryear{{Aly}}{{Aly}}{1994}]{1994PhRvE..49.3771A}
{Aly} J.~J.,  1994, \mn@doi [\pre] {10.1103/PhysRevE.49.3771}, \href
  {http://adsabs.harvard.edu/abs/1994PhRvE..49.3771A} {49, 3771}

\bibitem[\protect\citeauthoryear{{Arcoragi}, {Bonnell}, {Martel}, {Bastien}  \&
  {Benz}}{{Arcoragi} et~al.}{1991}]{1991ApJ...380..476A}
{Arcoragi} J.-P.,  {Bonnell} I.,  {Martel} H.,  {Bastien} P.,   {Benz} W.,
  1991, \mn@doi [\apj] {10.1086/170605}, \href
  {http://adsabs.harvard.edu/abs/1991ApJ...380..476A} {380, 476}

\bibitem[\protect\citeauthoryear{{Arzoumanian} et~al.,}{{Arzoumanian}
  et~al.}{2011}]{2011A&A...529L...6A}
{Arzoumanian} D.,  et~al., 2011, \mn@doi [\aap] {10.1051/0004-6361/201116596},
  \href {http://adsabs.harvard.edu/abs/2011A%26A...529L...6A} {529, L6}

\bibitem[\protect\citeauthoryear{{Aschwanden}}{{Aschwanden}}{2005}]{2005psci.book.....A}
{Aschwanden} M.~J.,  2005, {Physics of the Solar Corona. An Introduction with
  Problems and Solutions (2nd edition)}

\bibitem[\protect\citeauthoryear{{Binney} \& {Tremaine}}{{Binney} \&
  {Tremaine}}{2008}]{2008gady.book.....B}
{Binney} J.,  {Tremaine} S.,  2008, {Galactic Dynamics: Second Edition}.
Princeton University Press

\bibitem[\protect\citeauthoryear{{Breysse}, {Kamionkowski}  \&
  {Benson}}{{Breysse} et~al.}{2014}]{2014MNRAS.437.2675B}
{Breysse} P.~C.,  {Kamionkowski} M.,   {Benson} A.,  2014, \mn@doi [\mnras]
  {10.1093/mnras/stt2077}, \href
  {http://adsabs.harvard.edu/abs/2014MNRAS.437.2675B} {437, 2675}

\bibitem[\protect\citeauthoryear{{Bufferand}, {Ciraolo}, {Ghendrih}, {Tamain},
  {Bagnoli}, {Lepri}  \& {Livi}}{{Bufferand}
  et~al.}{2010}]{2010JPhCS.260a2005B}
{Bufferand} H.,  {Ciraolo} G.,  {Ghendrih} P.,  {Tamain} P.,  {Bagnoli} F.,
  {Lepri} S.,   {Livi} R.,  2010, \mn@doi [Journal of Physics Conference
  Series] {10.1088/1742-6596/260/1/012005}, \href
  {http://adsabs.harvard.edu/abs/2010JPhCS.260a2005B} {260, 012005}

\bibitem[\protect\citeauthoryear{{Bufferand}, {Ciraolo}, {Ghendrih}, {Lepri}
  \& {Livi}}{{Bufferand} et~al.}{2013}]{2013PhRvE..87b3102B}
{Bufferand} H.,  {Ciraolo} G.,  {Ghendrih} P.,  {Lepri} S.,   {Livi} R.,  2013,
  \mn@doi [\pre] {10.1103/PhysRevE.87.023102}, \href
  {http://adsabs.harvard.edu/abs/2013PhRvE..87b3102B} {87, 023102}

\bibitem[\protect\citeauthoryear{{Burkert} \& {Hartmann}}{{Burkert} \&
  {Hartmann}}{2004}]{2004ApJ...616..288B}
{Burkert} A.,  {Hartmann} L.,  2004, \mn@doi [\apj] {10.1086/424895}, \href
  {http://adsabs.harvard.edu/abs/2004ApJ...616..288B} {616, 288}

\bibitem[\protect\citeauthoryear{Campa, Dauxois, Fanelli  \& Ruffo}{Campa
  et~al.}{2014}]{CampaEtAl:book}
Campa A.,  Dauxois T.,  Fanelli D.,   Ruffo S.,  2014, Physics of Long-Range
  Interacting Systems.
Oxford University Press, Oxford

\bibitem[\protect\citeauthoryear{{Casetti}}{{Casetti}}{1995}]{1995PhyS...51...29C}
{Casetti} L.,  1995, \mn@doi [Physica Scripta] {10.1088/0031-8949/51/1/005},
  \href {http://adsabs.harvard.edu/abs/1995PhyS...51...29C} {51, 29}

\bibitem[\protect\citeauthoryear{{Casetti} \& {Gupta}}{{Casetti} \&
  {Gupta}}{2014}]{2014EPJB...87...91C}
{Casetti} L.,  {Gupta} S.,  2014, \mn@doi [European Physical Journal B]
  {10.1140/epjb/e2014-50136-y}, \href
  {http://adsabs.harvard.edu/abs/2014EPJB...87...91C} {87, 91}

\bibitem[\protect\citeauthoryear{{Chavanis}}{{Chavanis}}{2006}]{2006CRPhy...7..331C}
{Chavanis} P.-H.,  2006, \mn@doi [Comptes Rendus Physique]
  {10.1016/j.crhy.2006.01.005}, \href
  {http://adsabs.harvard.edu/abs/2006CRPhy...7..331C} {7, 331}

\bibitem[\protect\citeauthoryear{{Chira}, {Siebenmorgen}, {Henning}  \&
  {Kainulainen}}{{Chira} et~al.}{2016}]{2016A&A...592A..90C}
{Chira} R.-A.,  {Siebenmorgen} R.,  {Henning} T.,   {Kainulainen} J.,  2016,
  \mn@doi [\aap] {10.1051/0004-6361/201528028}, \href
  {http://adsabs.harvard.edu/abs/2016A%26A...592A..90C} {592, A90}

\bibitem[\protect\citeauthoryear{{Crutcher}}{{Crutcher}}{2012}]{2012ARA&A..50...29C}
{Crutcher} R.~M.,  2012, \mn@doi [\araa] {10.1146/annurev-astro-081811-125514},
  \href {http://adsabs.harvard.edu/abs/2012ARA%26A..50...29C} {50, 29}

\bibitem[\protect\citeauthoryear{{Cuperman}, {Goldstein}  \&
  {Lecar}}{{Cuperman} et~al.}{1969}]{1969MNRAS.146..161C}
{Cuperman} S.,  {Goldstein} S.,   {Lecar} M.,  1969, \mn@doi [\mnras]
  {10.1093/mnras/146.2.161}, \href
  {http://adsabs.harvard.edu/abs/1969MNRAS.146..161C} {146, 161}

\bibitem[\protect\citeauthoryear{{Danilov}}{{Danilov}}{1988}]{1988SvA....32..374D}
{Danilov} V.~M.,  1988, \sovast, \href
  {http://adsabs.harvard.edu/abs/1988SvA....32..374D} {32, 374}

\bibitem[\protect\citeauthoryear{{Davidson} \& {Qin}}{{Davidson} \&
  {Qin}}{2001}]{2001picp.book.....D}
{Davidson} R.~C.,  {Qin} H.,  2001, {Physics of Intense Charged Particle Beams
  in High Energy Accelerators}.
World Scientific Press, \mn@doi{10.1142/p250}

\bibitem[\protect\citeauthoryear{{Di Cintio} \& {Ciotti}}{{Di Cintio} \&
  {Ciotti}}{2011}]{2011IJBC...21.2279D}
{Di Cintio} P.,  {Ciotti} L.,  2011, \mn@doi [International Journal of
  Bifurcation and Chaos] {10.1142/S021812741102977X}, \href
  {http://adsabs.harvard.edu/abs/2011IJBC...21.2279D} {21, 2279}

\bibitem[\protect\citeauthoryear{{Di Cintio}, {Ciotti}  \& {Nipoti}}{{Di
  Cintio} et~al.}{2013}]{2013MNRAS.431.3177D}
{Di Cintio} P.,  {Ciotti} L.,   {Nipoti} C.,  2013, \mn@doi [\mnras]
  {10.1093/mnras/stt403}, \href
  {http://adsabs.harvard.edu/abs/2013MNRAS.431.3177D} {431, 3177}

\bibitem[\protect\citeauthoryear{{Di Cintio}, {Livi}, {Bufferand}, {Ciraolo},
  {Lepri}  \& {Straka}}{{Di Cintio} et~al.}{2015}]{2015PhRvE..92f2108D}
{Di Cintio} P.,  {Livi} R.,  {Bufferand} H.,  {Ciraolo} G.,  {Lepri} S.,
  {Straka} M.~J.,  2015, \mn@doi [\pre] {10.1103/PhysRevE.92.062108}, \href
  {http://adsabs.harvard.edu/abs/2015PhRvE..92f2108D} {92, 062108}

\bibitem[\protect\citeauthoryear{{Di Cintio}, {Livi}, {Lepri}  \&
  {Ciraolo}}{{Di Cintio} et~al.}{2017}]{2017PhRvE..95d3203D}
{Di Cintio} P.,  {Livi} R.,  {Lepri} S.,   {Ciraolo} G.,  2017, \mn@doi [\pre]
  {10.1103/PhysRevE.95.043203}, \href
  {http://adsabs.harvard.edu/abs/2017PhRvE..95d3203D} {95, 043203}

\bibitem[\protect\citeauthoryear{{Drukier}, {Cohn}, {Lugger}  \&
  {Yong}}{{Drukier} et~al.}{1999}]{1999ApJ...518..233D}
{Drukier} G.~A.,  {Cohn} H.~N.,  {Lugger} P.~M.,   {Yong} H.,  1999, \mn@doi
  [\apj] {10.1086/307243}, \href
  {http://adsabs.harvard.edu/abs/1999ApJ...518..233D} {518, 233}

\bibitem[\protect\citeauthoryear{{Federrath}}{{Federrath}}{2016}]{2016MNRAS.457..375F}
{Federrath} C.,  2016, \mn@doi [\mnras] {10.1093/mnras/stv2880}, \href
  {http://adsabs.harvard.edu/abs/2016MNRAS.457..375F} {457, 375}

\bibitem[\protect\citeauthoryear{{Fellhauer}, {Kroupa}, {Baumgardt}, {Bien},
  {Boily}, {Spurzem}  \& {Wassmer}}{{Fellhauer}
  et~al.}{2000}]{2000NewA....5..305F}
{Fellhauer} M.,  {Kroupa} P.,  {Baumgardt} H.,  {Bien} R.,  {Boily} C.~M.,
  {Spurzem} R.,   {Wassmer} N.,  2000, \mn@doi [\na]
  {10.1016/S1384-1076(00)00032-4}, \href
  {http://adsabs.harvard.edu/abs/2000NewA....5..305F} {5, 305}

\bibitem[\protect\citeauthoryear{{Gabrielli}, {Joyce}  \& {Marcos}}{{Gabrielli}
  et~al.}{2010}]{GabrielliJoyceMarcos:prl2010}
{Gabrielli} A.,  {Joyce} M.,   {Marcos} B.,  2010, \mn@doi [Physical Review
  Letters] {10.1103/PhysRevLett.105.210602}, \href
  {http://adsabs.harvard.edu/abs/2010PhRvL.105u0602G} {105, 210602}

\bibitem[\protect\citeauthoryear{{Gehman}, {Adams}  \& {Watkins}}{{Gehman}
  et~al.}{1996}]{1996ApJ...472..673G}
{Gehman} C.~S.,  {Adams} F.~C.,   {Watkins} R.,  1996, \mn@doi [\apj]
  {10.1086/178098}, \href {http://adsabs.harvard.edu/abs/1996ApJ...472..673G}
  {472, 673}

\bibitem[\protect\citeauthoryear{Golub \& Pasachoff}{Golub \&
  Pasachoff}{2009}]{GolubPasachoff:book}
Golub L.,  Pasachoff J.~M.,  2009, The Solar Corona, second edn.
Cambridge University Press, Cambridge

\bibitem[\protect\citeauthoryear{{G{\'o}mez} \&
  {V{\'a}zquez-Semadeni}}{{G{\'o}mez} \&
  {V{\'a}zquez-Semadeni}}{2014}]{2014ApJ...791..124G}
{G{\'o}mez} G.~C.,  {V{\'a}zquez-Semadeni} E.,  2014, \mn@doi [\apj]
  {10.1088/0004-637X/791/2/124}, \href
  {http://adsabs.harvard.edu/abs/2014ApJ...791..124G} {791, 124}

\bibitem[\protect\citeauthoryear{{Gompper}, {Ihle}, {Kroll}  \&
  {Winkler}}{{Gompper} et~al.}{2009}]{2009acsa.book....1G}
{Gompper} G.,  {Ihle} T.,  {Kroll} D.~M.,   {Winkler} R.~G.,  2009,
  {Multi-Particle Collision Dynamics: A Particle-Based Mesoscale Simulation
  Approach to the Hydrodynamics of Complex Fluids}.
p.~1, \mn@doi{10.1007/978-3-540-87706-6_1}

\bibitem[\protect\citeauthoryear{{Goodman}, {Barranco}, {Wilner}  \&
  {Heyer}}{{Goodman} et~al.}{1998}]{0004-637X-504-1-223}
{Goodman} A.~A.,  {Barranco} J.~A.,  {Wilner} D.~J.,   {Heyer} M.~H.,  1998,
  \apj, 504, 223

\bibitem[\protect\citeauthoryear{{Gupta} \& {Casetti}}{{Gupta} \&
  {Casetti}}{2016}]{njp2016}
{Gupta} S.,  {Casetti} L.,  2016, \mn@doi [New Journal of Physics]
  {10.1088/1367-2630/18/10/103051}, \href
  {http://adsabs.harvard.edu/abs/2016NJPh...18j3051G} {18, 103051}

\bibitem[\protect\citeauthoryear{{Hockney} \& {Eastwood}}{{Hockney} \&
  {Eastwood}}{1981}]{1981csup.book.....H}
{Hockney} R.~W.,  {Eastwood} J.~W.,  1981, {Computer Simulation Using
  Particles}

\bibitem[\protect\citeauthoryear{{Holloway}, {Fiorito}, {Shkvarunets},
  {O'Shea}, {Benson}, {Douglas}, {Evtushenko}  \& {Jordan}}{{Holloway}
  et~al.}{2008}]{2008PhRvS..11h2801H}
{Holloway} M.~A.,  {Fiorito} R.~B.,  {Shkvarunets} A.~G.,  {O'Shea} P.~G.,
  {Benson} S.~V.,  {Douglas} D.,  {Evtushenko} P.,   {Jordan} K.,  2008,
  \mn@doi [Physical Review Special Topics Accelerators and Beams]
  {10.1103/PhysRevSTAB.11.082801}, \href
  {http://adsabs.harvard.edu/abs/2008PhRvS..11h2801H} {11, 082801}

\bibitem[\protect\citeauthoryear{{Kapral}}{{Kapral}}{2008}]{kapral08}
{Kapral} R.,  2008, {Multiparticle Collision Dynamics: Simulation of Complex
  Systems on Mesoscales}.
pp 89--146, \mn@doi{10.1002/9780470371572.ch2}

\bibitem[\protect\citeauthoryear{{Katz} \& {Lynden-Bell}}{{Katz} \&
  {Lynden-Bell}}{1978}]{1978MNRAS.184..709K}
{Katz} J.,  {Lynden-Bell} D.,  1978, \mn@doi [\mnras]
  {10.1093/mnras/184.4.709}, \href
  {http://adsabs.harvard.edu/abs/1978MNRAS.184..709K} {184, 709}

\bibitem[\protect\citeauthoryear{{Kinoshita}, {Yoshida}  \&
  {Nakai}}{{Kinoshita} et~al.}{1991}]{1991CeMDA..50...59K}
{Kinoshita} H.,  {Yoshida} H.,   {Nakai} H.,  1991, Celestial Mechanics and
  Dynamical Astronomy, \href
  {http://adsabs.harvard.edu/abs/1991CeMDA..50...59K} {50, 59}

\bibitem[\protect\citeauthoryear{{Klessen}}{{Klessen}}{2011}]{2011EAS....51..133K}
{Klessen} R.~S.,  2011, in {Charbonnel} C.,  {Montmerle} T.,  eds,  EAS
  Publications Series Vol. 51, EAS Publications Series. pp 133--167 (\mn@eprint
  {arXiv} {1109.0467}), \mn@doi{10.1051/eas/1151009}

\bibitem[\protect\citeauthoryear{{Levin}, {Pakter}  \& {Teles}}{{Levin}
  et~al.}{2008}]{2008PhRvL.100d0604L}
{Levin} Y.,  {Pakter} R.,   {Teles} T.~N.,  2008, \mn@doi [Physical Review
  Letters] {10.1103/PhysRevLett.100.040604}, \href
  {http://adsabs.harvard.edu/abs/2008PhRvL.100d0604L} {100, 040604}

\bibitem[\protect\citeauthoryear{Levin, Pakter, Rizzato, Teles  \&
  Benetti}{Levin et~al.}{2014}]{LevinEtAlphysrep:2014}
Levin Y.,  Pakter R.,  Rizzato F.~B.,  Teles T.~N.,   Benetti F. P.~C.,  2014,
  \mn@doi [Physics Reports] {http://dx.doi.org/10.1016/j.physrep.2013.10.001},
  535, 1

\bibitem[\protect\citeauthoryear{{Li}, {Fang}, {Henning}  \&
  {Kainulainen}}{{Li} et~al.}{2013}]{2013MNRAS.436.3707L}
{Li} H.-b.,  {Fang} M.,  {Henning} T.,   {Kainulainen} J.,  2013, \mn@doi
  [\mnras] {10.1093/mnras/stt1849}, \href
  {http://adsabs.harvard.edu/abs/2013MNRAS.436.3707L} {436, 3707}

\bibitem[\protect\citeauthoryear{{Lou}}{{Lou}}{2015}]{2015MNRAS.454.2815L}
{Lou} Y.-Q.,  2015, \mn@doi [\mnras] {10.1093/mnras/stv1912}, \href
  {http://adsabs.harvard.edu/abs/2015MNRAS.454.2815L} {454, 2815}

\bibitem[\protect\citeauthoryear{{Loubser}, {Sansom},
  {S{\'a}nchez-Bl{\'a}zquez}, {Soechting}  \& {Bromage}}{{Loubser}
  et~al.}{2008}]{LoubserEtAl:mnras2008}
{Loubser} S.~I.,  {Sansom} A.~E.,  {S{\'a}nchez-Bl{\'a}zquez} P.,  {Soechting}
  I.~K.,   {Bromage} G.~E.,  2008, \mn@doi [\mnras]
  {10.1111/j.1365-2966.2008.13813.x}, \href
  {http://adsabs.harvard.edu/abs/2008MNRAS.391.1009L} {391, 1009}

\bibitem[\protect\citeauthoryear{{Malevanets} \& {Kapral}}{{Malevanets} \&
  {Kapral}}{1999}]{1999JChPh.110.8605M}
{Malevanets} A.,  {Kapral} R.,  1999, \mn@doi [\jcp] {10.1063/1.478857}, \href
  {http://adsabs.harvard.edu/abs/1999JChPh.110.8605M} {110, 8605}

\bibitem[\protect\citeauthoryear{{Malevanets} \& {Kapral}}{{Malevanets} \&
  {Kapral}}{2004}]{2004LNP...640..116M}
{Malevanets} A.,  {Kapral} R.,  2004, in {Karttunen} M.,  {Lukkarinen} A.,
  {Vattulainen} I.,  eds,  Lecture Notes in Physics, Berlin Springer Verlag
  Vol. 640, Novel Methods in Soft Matter Simulations. pp 116--149,
  \mn@doi{10.1007/b95265}

\bibitem[\protect\citeauthoryear{{Marcos}}{{Marcos}}{2013}]{Marcos:pre2013}
{Marcos} B.,  2013, \mn@doi [\pre] {10.1103/PhysRevE.88.032112}, \href
  {http://adsabs.harvard.edu/abs/2013PhRvE..88c2112M} {88, 032112}

\bibitem[\protect\citeauthoryear{{Marcos}, {Gabrielli}  \& {Joyce}}{{Marcos}
  et~al.}{2017}]{MarcosGabrielliJoyce:pre2017}
{Marcos} B.,  {Gabrielli} A.,   {Joyce} M.,  2017, \mn@doi [\pre]
  {10.1103/PhysRevE.96.032102}, \href
  {http://adsabs.harvard.edu/abs/2017PhRvE..96c2102M} {96, 032102}

\bibitem[\protect\citeauthoryear{{Meyer-Vernet}, {Hoang}  \&
  {Moncuquet}}{{Meyer-Vernet} et~al.}{1993}]{1993JGR....9821163M}
{Meyer-Vernet} N.,  {Hoang} S.,   {Moncuquet} M.,  1993, \mn@doi [\jgr]
  {10.1029/93JA02587}, \href
  {http://adsabs.harvard.edu/abs/1993JGR....9821163M} {98, 21}

\bibitem[\protect\citeauthoryear{{Meyer-Vernet}, {Moncuquet}  \&
  {Hoang}}{{Meyer-Vernet} et~al.}{1995}]{MeyerVernet1995202}
{Meyer-Vernet} N.,  {Moncuquet} M.,   {Hoang} S.,  1995, \mn@doi [Icarus]
  {http://dx.doi.org/10.1006/icar.1995.1121}, 116, 202

\bibitem[\protect\citeauthoryear{{Myers}}{{Myers}}{2015}]{2015ApJ...806..226M}
{Myers} P.~C.,  2015, \mn@doi [\apj] {10.1088/0004-637X/806/2/226}, \href
  {http://adsabs.harvard.edu/abs/2015ApJ...806..226M} {806, 226}

\bibitem[\protect\citeauthoryear{{Nghiem}, {Chauvin}, {Simeoni}  \&
  {Uriot}}{{Nghiem} et~al.}{2014}]{2014ApPhL.104g4109N}
{Nghiem} P.~A.~P.,  {Chauvin} N.,  {Simeoni} W.,   {Uriot} D.,  2014, \mn@doi
  [Applied Physics Letters] {10.1063/1.4866400}, \href
  {http://adsabs.harvard.edu/abs/2014ApPhL.104g4109N} {104, 074109}

\bibitem[\protect\citeauthoryear{{Ostriker}}{{Ostriker}}{1964a}]{1964ApJ...140.1056O}
{Ostriker} J.,  1964a, \mn@doi [\apj] {10.1086/148005}, \href
  {http://adsabs.harvard.edu/abs/1964ApJ...140.1056O} {140, 1056}

\bibitem[\protect\citeauthoryear{{Ostriker}}{{Ostriker}}{1964b}]{1964ApJ...140.1529O}
{Ostriker} J.,  1964b, \mn@doi [\apj] {10.1086/148057}, \href
  {http://adsabs.harvard.edu/abs/1964ApJ...140.1529O} {140, 1529}

\bibitem[\protect\citeauthoryear{{Pineda}, {Goodman}, {Arce}, {Caselli},
  {Foster}, {Myers}  \& {Rosolowsky}}{{Pineda}
  et~al.}{2010}]{2010ApJ...712L.116P}
{Pineda} J.~E.,  {Goodman} A.~A.,  {Arce} H.~G.,  {Caselli} P.,  {Foster}
  J.~B.,  {Myers} P.~C.,   {Rosolowsky} E.~W.,  2010, \mn@doi [\apjl]
  {10.1088/2041-8205/712/1/L116}, \href
  {http://adsabs.harvard.edu/abs/2010ApJ...712L.116P} {712, L116}

\bibitem[\protect\citeauthoryear{{Recchi}, {Hacar}  \& {Palestini}}{{Recchi}
  et~al.}{2013}]{2013A&A...558A..27R}
{Recchi} S.,  {Hacar} A.,   {Palestini} A.,  2013, \mn@doi [\aap]
  {10.1051/0004-6361/201321565}, \href
  {http://adsabs.harvard.edu/abs/2013A%26A...558A..27R} {558, A27}

\bibitem[\protect\citeauthoryear{{Recchi}, {Hacar}  \& {Palestini}}{{Recchi}
  et~al.}{2014}]{2014MNRAS.444.1775R}
{Recchi} S.,  {Hacar} A.,   {Palestini} A.,  2014, \mn@doi [\mnras]
  {10.1093/mnras/stu1566}, \href
  {http://adsabs.harvard.edu/abs/2014MNRAS.444.1775R} {444, 1775}

\bibitem[\protect\citeauthoryear{{Riabko}, {Ellison}, {Kang}, {Lee}, {Li},
  {Liu}, {Pei}  \& {Wang}}{{Riabko} et~al.}{1995}]{1995PhRvE..51.3529R}
{Riabko} A.,  {Ellison} M.,  {Kang} X.,  {Lee} S.~Y.,  {Li} D.,  {Liu} J.~Y.,
  {Pei} X.,   {Wang} L.,  1995, \mn@doi [\pre] {10.1103/PhysRevE.51.3529},
  \href {http://adsabs.harvard.edu/abs/1995PhRvE..51.3529R} {51, 3529}

\bibitem[\protect\citeauthoryear{{Rouleau} \& {Bastien}}{{Rouleau} \&
  {Bastien}}{1990}]{1990ApJ...355..172R}
{Rouleau} F.,  {Bastien} P.,  1990, \mn@doi [\apj] {10.1086/168751}, \href
  {http://adsabs.harvard.edu/abs/1990ApJ...355..172R} {355, 172}

\bibitem[\protect\citeauthoryear{{Ruffini}, {Arg{\"u}elles}  \&
  {Rueda}}{{Ruffini} et~al.}{2015}]{2015MNRAS.451..622R}
{Ruffini} R.,  {Arg{\"u}elles} C.~R.,   {Rueda} J.~A.,  2015, \mn@doi [\mnras]
  {10.1093/mnras/stv1016}, \href
  {http://adsabs.harvard.edu/abs/2015MNRAS.451..622R} {451, 622}

\bibitem[\protect\citeauthoryear{{Ryder}}{{Ryder}}{2005}]{tesiryder}
{Ryder} J.,  2005, PhD thesis, Oxford University, UK

\bibitem[\protect\citeauthoryear{{Saur}, {Neubauer}, {Connerney}, {Zarka}  \&
  {Kivelson}}{{Saur} et~al.}{2004}]{2004jpsm.book..537S}
{Saur} J.,  {Neubauer} F.~M.,  {Connerney} J.~E.~P.,  {Zarka} P.,   {Kivelson}
  M.~G.,  2004, {Plasma interaction of Io with its plasma torus}.
pp 537--560

\bibitem[\protect\citeauthoryear{{Schulz}, {Dehnen}, {Jungman}  \&
  {Tremaine}}{{Schulz} et~al.}{2013}]{2013MNRAS.431...49S}
{Schulz} A.~E.,  {Dehnen} W.,  {Jungman} G.,   {Tremaine} S.,  2013, \mn@doi
  [\mnras] {10.1093/mnras/stt073}, \href
  {http://adsabs.harvard.edu/abs/2013MNRAS.431...49S} {431, 49}

\bibitem[\protect\citeauthoryear{{Scudder}}{{Scudder}}{1992a}]{1992ApJ...398..299S}
{Scudder} J.~D.,  1992a, \mn@doi [\apj] {10.1086/171858}, \href
  {http://adsabs.harvard.edu/abs/1992ApJ...398..299S} {398, 299}

\bibitem[\protect\citeauthoryear{{Scudder}}{{Scudder}}{1992b}]{1992ApJ...398..319S}
{Scudder} J.~D.,  1992b, \mn@doi [\apj] {10.1086/171859}, \href
  {http://adsabs.harvard.edu/abs/1992ApJ...398..319S} {398, 319}

\bibitem[\protect\citeauthoryear{{Shadmehri}}{{Shadmehri}}{2005}]{2005MNRAS.356.1429S}
{Shadmehri} M.,  2005, \mn@doi [\mnras] {10.1111/j.1365-2966.2004.08575.x},
  \href {http://adsabs.harvard.edu/abs/2005MNRAS.356.1429S} {356, 1429}

\bibitem[\protect\citeauthoryear{{Silvestre} \& {Rocha Filho}}{{Silvestre} \&
  {Rocha Filho}}{2016}]{2016PhLA..380..337S}
{Silvestre} C.~H.,  {Rocha Filho} T.~M.,  2016, \mn@doi [Physics Letters A]
  {10.1016/j.physleta.2015.10.042}, \href
  {http://adsabs.harvard.edu/abs/2016PhLA..380..337S} {380, 337}

\bibitem[\protect\citeauthoryear{{Stod{\'o}lkiewicz}}{{Stod{\'o}lkiewicz}}{1963}]{1963AcA....13...30S}
{Stod{\'o}lkiewicz} J.~S.,  1963, \actaa, \href
  {http://adsabs.harvard.edu/abs/1963AcA....13...30S} {13, 30}

\bibitem[\protect\citeauthoryear{{Teles}, {Levin}, {Pakter}  \&
  {Rizzato}}{{Teles} et~al.}{2010}]{TelesLevinPakterRizzato:jstat2010}
{Teles} T.~N.,  {Levin} Y.,  {Pakter} R.,   {Rizzato} F.~B.,  2010, \mn@doi
  [Journal of Statistical Mechanics: Theory and Experiment]
  {10.1088/1742-5468/2010/05/P05007}, \href
  {http://adsabs.harvard.edu/abs/2010JSMTE..05..007T} {5, 05007}

\bibitem[\protect\citeauthoryear{{Teles}, {Levin}  \& {Pakter}}{{Teles}
  et~al.}{2011}]{2011MNRAS.417L..21T}
{Teles} T.~N.,  {Levin} Y.,   {Pakter} R.,  2011, \mn@doi [\mnras]
  {10.1111/j.1745-3933.2011.01112.x}, \href
  {http://adsabs.harvard.edu/abs/2011MNRAS.417L..21T} {417, L21}

\bibitem[\protect\citeauthoryear{{Teles}, {Gupta}, {Di Cintio}  \&
  {Casetti}}{{Teles} et~al.}{2015}]{2015PhRvE..92b0101T}
{Teles} T.~N.,  {Gupta} S.,  {Di Cintio} P.,   {Casetti} L.,  2015, \mn@doi
  [\pre] {10.1103/PhysRevE.92.020101}, \href
  {http://adsabs.harvard.edu/abs/2015PhRvE..92b0101T} {92, 020101}

\bibitem[\protect\citeauthoryear{{Teles}, {Gupta}, {Di Cintio}  \&
  {Casetti}}{{Teles} et~al.}{2016}]{2016PhRvE..93f6102T}
{Teles} T.~N.,  {Gupta} S.,  {Di Cintio} P.,   {Casetti} L.,  2016, \mn@doi
  [\pre] {10.1103/PhysRevE.93.066102}, \href
  {http://adsabs.harvard.edu/abs/2016PhRvE..93f6102T} {93, 066102}

\bibitem[\protect\citeauthoryear{{Toci} \& {Galli}}{{Toci} \&
  {Galli}}{2015a}]{2015MNRAS.446.2110T}
{Toci} C.,  {Galli} D.,  2015a, \mn@doi [\mnras] {10.1093/mnras/stu2168}, \href
  {http://adsabs.harvard.edu/abs/2015MNRAS.446.2110T} {446, 2110}

\bibitem[\protect\citeauthoryear{{Toci} \& {Galli}}{{Toci} \&
  {Galli}}{2015b}]{2015MNRAS.446.2118T}
{Toci} C.,  {Galli} D.,  2015b, \mn@doi [\mnras] {10.1093/mnras/stu2194}, \href
  {http://adsabs.harvard.edu/abs/2015MNRAS.446.2118T} {446, 2118}

\bibitem[\protect\citeauthoryear{{Trenti} \& {Bertin}}{{Trenti} \&
  {Bertin}}{2006}]{TrentiBertin:Apj2006}
{Trenti} M.,  {Bertin} G.,  2006, \mn@doi [\apj] {10.1086/498637}, \href
  {http://adsabs.harvard.edu/abs/2006ApJ...637..717T} {637, 717}

\bibitem[\protect\citeauthoryear{{Veale}, {Ma}, {Greene}, {Thomas},
  {Blakeslee}, {Walsh}  \& {Ito}}{{Veale} et~al.}{2017}]{2017arXiv170800870V}
{Veale} M.,  {Ma} C.-P.,  {Greene} J.~E.,  {Thomas} J.,  {Blakeslee} J.~P.,
  {Walsh} J.~L.,   {Ito} J.,  2017, preprint, \href
  {http://adsabs.harvard.edu/abs/2017arXiv170800870V} {} (\mn@eprint {arXiv}
  {1708.00870})

\bibitem[\protect\citeauthoryear{{Wangler}, {Crandall}, {Ryne}  \&
  {Wang}}{{Wangler} et~al.}{1998}]{1998PhRvS...1h4201W}
{Wangler} T.~P.,  {Crandall} K.~R.,  {Ryne} R.,   {Wang} T.~S.,  1998, \mn@doi
  [Physical Review Special Topics Accelerators and Beams]
  {10.1103/PhysRevSTAB.1.084201}, \href
  {http://adsabs.harvard.edu/abs/1998PhRvS...1h4201W} {1, 084201}

\bibitem[\protect\citeauthoryear{Wise, McNamara  \& Murray}{Wise
  et~al.}{2004}]{0004-637X-601-1-184}
Wise M.~W.,  McNamara B.~R.,   Murray S.~S.,  2004, \apj, 601, 184

\bibitem[\protect\citeauthoryear{Worrakitpoonpon}{Worrakitpoonpon}{2011}]{W:thesis}
Worrakitpoonpon T.,  2011, PhD thesis, Universit\'e Paris VI

\bibitem[\protect\citeauthoryear{{Xiong}, {Chen}, {Yang}, {Fang}, {Zhang},
  {Zhang}, {Du}  \& {Long}}{{Xiong} et~al.}{2017}]{2017arXiv170301394X}
{Xiong} F.,  {Chen} X.,  {Yang} J.,  {Fang} M.,  {Zhang} S.,  {Zhang} M.,  {Du}
  X.,   {Long} W.,  2017, preprint, \href
  {http://adsabs.harvard.edu/abs/2017arXiv170301394X} {} (\mn@eprint {arXiv}
  {1703.01394})

\bibitem[\protect\citeauthoryear{{van Albada}}{{van
  Albada}}{1982}]{vanAlbada:mnras1982}
{van Albada} T.~S.,  1982, \mn@doi [\mnras] {10.1093/mnras/201.4.939}, \href
  {http://adsabs.harvard.edu/abs/1982MNRAS.201..939V} {201, 939}

\makeatother
\end{thebibliography}
\end{document}